\documentclass{emulateapj}
\usepackage{graphicx}

\begin{document}

\def\etal{et al.\ \rm}

\title{A Physical Bias in Cosmological Simulations}

\author{Weike Xiao\altaffilmark{1}, Zhengfan Sun,and Heng Hao}

\affil{Center for Astrophysics, University of Science and
Technology of China,\\ Hefei, Anhui, 230026,China}

\email{wkxiao@mail.ustc.edu.cn}


\begin{abstract}
Numerical simulations play an important role in the study of the
structure formation of the universe. However, the mass resolution
in current simulations is still poor. Due to technical
difficulties, it is necessary to use both a greatly reduced number
density of particles and a greatly raised unit particle mass.
Consequently, the particle masses used in cosmological simulations
are about $10^{70}$ times larger than the Gev candidates in
particle physics. This is a huge physical bias that cannot be
neglected when interpreting the results of the simulations. Here
we discuss how such a bias affects cold dark matter (CDM)
cosmological simulations. We find that the small-scale properties
of the CDM particle system are changed in two aspects. (1) An
upper limit is imposed on the spatial resolution of the simulation
results. (2) Most importantly, an unexpected short mean free path
is produced, and the corresponding two-body scattering cross
section is close to the value expected in the self-interaction
dark matter model. Since the mean free path of real CDM particle
systems is much longer than that in the simulations, our results
imply that (1) there is probably no "cusp problem" in real CDM
halo, and (2) a much longer time is needed to form new virialized
halos in real CDM particle systems than in the simulations. The
last result can help us understand the 'substructure problem'. Our
discussion can also explain why the massive halos in the
simulations may have smaller concentration coefficients.

\end{abstract}



\section{Introduction.}
\label{sect:intro}

Thanks to the rapid development of numerical computation
technology, cosmological $N$-body simulations have become an
increasingly important tool in our understanding of the
gravitational clustering process in the universe. Suto (2003)
noted that the number of particles used in simulations has risen
exponentially with time, from about 400 in 1975 to about $10^9$
today. This means that current $N$-body simulations are available
to give reliable results on some complex cosmological issues such
as large-scale structure formation (Einasto,2001) and the
properties of dark matter halos (Cooray\&
Sheth,2002;Navarro,Frenk\& White,1996). Sometimes such numerical
methods are the {\it only} way to get a qualitative picture of the
system for which there is no mature theory. This is especially the
case when strong nonlinear evolution on small scales is involved,
such as in the central region of dark matter halos (Primack,2003)
and in the processes of galaxy formation and merging
(Kauffmann,White\& Guiderdoni,1993), etc.

High-resolution simulations based on the $\Lambda\!$CDM cosmology
model have shown nice agreement with the large-scale structure
observations in several cases (Freedman,W.L.\& Turner,M.S. 2003).
The predicted power spectrum in the simulations agrees well with
recent redshift surveys like the Sloan Digital Sky Survey (Tegmark
et al.,2003) and the Two-Degree Field Galaxy Redshift Survey
(Percival,W.J. 2001). But the simulations based on cold dark
matter(CDM) models do not work so well on small scales. One aspect
is that the central density profiles of halos predicted in these
simulations are too cuspy (Reed D,et al. 2003) to be consistent
with the observed results of dwarf galaxies and low surface
brightness galaxies (Primack,2003) or to be consistent with the
X-ray and gravitational lensing study of the distribution of dark
matter(DM) in galaxy clusters (Haruyoshi \& Kiyoshi 2004; Tyson et
al. 1998). Another aspect is that 10 times more substructures are
predicted than observed (Willman et al, 2002); this is called the
"substructure problem" or "dwarf galaxy problem"
(Primack,2003;Klypin,Kravtsov\& Valenzuela,1999). Some believe
that these problems come from deficiencies in the CDM model and
therefore introduce some alternative models, such as the
self-interaction dark matter(SIDM) model, warm dark matter (WDM)
model, etc. (Martin\& Jesper,2003;Celine,et
al.,2002;Spergel\&Steinhardt,2000)

Here we find that studying a common difficulty in current $N$-body
simulations will help us understand the above problems. Due to the
technical difficulties, the total number of particles $N$ we can
use even in recent simulations is still somewhat limited. The most
powerful 'high-resolution' simulation today with the largest total
number contains only about $512^3\-10^8$ particles. But at the
same time, we have to keep the mean density of the universe at a
preset level. So the mass of each particle in these simulations
has to be as high as about $10^{10} M_\odot$ (for a large-scale
simulation), comparable to the mass of a small galaxy. Now, when
we study the large-scale structure formation process, it is an
acceptable assumption that each particle in the simulation
represents one galaxy. However, when we study the small-scale
properties of dark matter particle systems, such as the CDM halo
density profile, the merging process of halos and the formation of
galaxies, the limitations arising from an inadequate $N$ will
emerge. For example, the number of particles in one CDM halo
($\sim 10^{13} M_\odot $) in current simulations ranges from
$10^2$ to $10^7$, and so the mass of each particle is greater than
$10^6 M_\odot$, therefore, what we can learn from such simulations
will concern halos made up of such "huge mass particles".

However, in the view of particle physicists, it is hard to accept
that the mass of a dark matter particle is some $10^{65}$ times
larger than that of usual GeV candidates. And it is also hard for
astrophysicists to accept, because a point particle with such a
huge mass is somehow a massive black hole, which will affect the
galaxy seriously. If we do not believe the mass of the CDM
particle should be the size of a small galaxy, then the particles
in the simulations are something different from real CDM
particles. In other words, we have unavoidably introduced a
physical bias in current simulations by having to simultaneously
reduce the particle number density $n$ and raise the mass of each
particle $m$ by a huge factor $f_{AE}$. The bias is like trying to
study the clustering properties of a huge society of ants by
simulating a colony of elephants: we call it the "ant-elephant
(AE) bias". Because CDM particle systems with AE bias can have
different dynamical and statistical properties, the clustering and
evolution scenario of CDM halos that we get from the simulations
may be very different from the truth. One might ask, since such a
physical bias is present in the current cosmological simulations,
are the simulation results reliable? To what extent will these
results reflect the actual situation? Is there any connection
between the bias and the "cusp problem" and the "substructure
problem" mentioned above? Learning how the AE bias affects the
simulation results and identifying the circumstances when the
effect is serious can help us reconstruct a correct scenario of
dark matter clustering and evolution from the simulation results.

When the factor $f_{AE}$ is less than about $10^4$, it is possible
to test the effects of the AE bias in the numerical simulation
directly. The number of particles used has already increased about
$10^6$ times in the past 30 years, which means the factor $f_{AE}$
has already decreased by a large amount, yet no serious changes in
large-scale properties has been reported.The factor $f_{AE}$ can
be as big as $10^{70}$ in the simulations, and therefore it is
indeed necessary to study how the simulation results will be
affected by such a physical bias.



\section{Physical difference caused by the AE bias}
\label{sect:application}


In this Letter, we study the physical difference the AE bias makes
on the cold dark matter systems, and find that the simulation
results are affected in two aspects.


\subsection{Constraint on the Spatial Resolution of Simulation Results}
\label{subsect:1_AU}

It is easy to see that the AE bias can impose an upper limit
$L_{min}$ on the spatial resolution of the simulation results. The
reason is that the AE bias has greatly reduced the particle number
density. To get reliable statistical results from simulation data
and in order to include enough simulations particles for
consideration the size of the statistical sample boxes cannot be
too small. For example, when discussing the galaxy formation
process at the center region of a CDM halo, one needs to calculate
the local density distribution $\rho$ of the CDM particles. Yet,
when estimating the local density at a certain position, the size
of the sample box $L$ can not be too small. An very small sample
box will include very limited simulation particles, and the
statistical result will be unreliable.

Consider an extreme case: if we use an ultra-small sample box with
volume $V_1 \sim m_{sim}/\rho$, then we can only find about $one$
simulation particle in it. Here the one simulation particle with
huge mass $m_{sim}$ actually represents all the real CDM particles
within the sample box. This is not a good approximation but it is
unavoidable because of technical difficulties, i.e. we are unable
to deal with so many particles. Consequently, the simulation
results can-not provide reliable information when the physical
processes occur on scales smaller than $L_{min1}\sim V_1^{1/3}$.
We have to be aware of this upper limit $L_{min1}$ when using the
simulation results to discuss the galaxy formation process and the
merging process of substructures.

$L_{min1}$ can be estimated in this way: the simulation particle
number within a sample box with size $L$ is about $N \simeq\rho
L^3 /m_{sim}$, where $m_{sim}$ is the mass of each simulation
particle and $\rho$ is the local density; set $N=1$ and we get
\begin{eqnarray}
 L_{min1}\nonumber &\simeq
 &\left(\frac{m_{sim}}{\rho}\right)^{\frac{1}{3}}\\
 & \simeq & \left(\frac{m_{sim}}{100 M_\odot}\right)^{1\over 3}\!\left(
 \frac{\rho}{\rho_{cr}}
 \frac{0.23}{\Omega_{cdm}}\frac{0.71}{h}\right)^{-{1\over 3}}(kpc/h)
\end{eqnarray}

Here we represent $m_{sim}$ and $\rho$ with the solar mass(
$M_\odot$) and the critical density of the universe
$\rho_{cr}=3H_0^2/8\pi G=1.88\times10^{-26}\cdot h^2\cdot
kg\cdot$. Equation(1) gives a physical limitation on the spacial
resolution of the simulation results.

However, it is still not enough, for it is hard to do statistical
analysis using a sample box that can include only about one
particle. In general, when we discuss the statistical results of a
given simulation sample,the statistical fluctuation must be taken
into account. For example, when we calculate the local density
$\rho$ at the halo center from the simulation result, the volume
of the sample box must be big enough to be able to contain
adequate simulation particles. The least particle number that one
sample box must contain depends on the requirement of a reliable
statistical result. A sample box with a size round $L_{min1}$ is
still not big enough and it can hardly give a reliable statistical
result of $\rho$.

Let us consider the statistical result of $\rho$ being
proportional to the particle number $N$ found in the sample box
with given volume $V$: then the statistical fluctuation of $\rho$
is about $\delta \rho/\rho \propto \delta N/N$. For a Gauss random
field, ${\delta N}/{N} \sim {1}/{\sqrt {N}}$. Suggest a reliable
statistical result requires $\delta \rho /\rho \propto \delta N /
N \sim 1/\sqrt{N} < a \%$, where $N$ is the particle number found
in the sample box with given volume $V$: then we get $N>10^4/a^2$.
This result means that the sample box must include at least
$N>10^4/a^2$ particles in order to give a reliable statistical
result of $\rho$. Thus, we get another limitation on the size of
the sample box $L_{min2}=V^{1/3}>(N V_1 )^{1\over 3}\simeq
L_{min1}({10^4}/{a^2})^{1\over3}$.

$L_{min2}$ gives a statistical limitation on the spatial
resolution of the simulation results. Here the factor $a$ varies
between $0$ and $100$ based on different problems: $L_{min2}$ is
commonly larger than $L_{min1}$. Only when $a\!\rightarrow\!100$
is $L_{min2}\!\rightarrow\!L_{min1}$. On scales smaller than
$L_{min2}$, not only can the statistical error be serious but the
physical process can be affected as well. Let us take the study of
the galaxy formation process at the halo center region as an
example again; one can find few massive particles in simulation
using a sample box smaller than $L_{min2}$, but one can find
numerous real CDM particles that can be smoothly distributed in
this field. Therefore the simulation of the CDM particle system
will offer much more asymmetry and instability to the
gravitational background for the study the galaxy formation
process on such a scale.

$L_{min1}$ and $L_{min2}$ can help us estimate a lower bound to
the scale of applicability of the simulation results. For example,
let us suggest that we have a $\Lambda\!$CDM cosmology simulation
with $m_{sim}=10^8 M_\odot$. For the center region of one CDM
halo, $\rho/\rho_{cr}\approx10^6 $, set $a=1$ and we get
$L_{min1}\approx1kpc/h$, $L_{min2}>10kpc/h$. When using these
simulation results to discuss the central properties of the halo,
such as the density profile or the galaxy formation process, the
AE bias must be taken into account. This is not an exciting
conclusion, and we have to accept the fact that by raising the
mass resolution $10^3$ times, the spatial resolution will only
increase by a factor of 10.

This AE bias effect comes from the statistical error $\delta N/N$,
which only occurs on small scales and can be neglected on large
scales where plenty of simulation particles are considered to
reach a reliable statistical result.


\subsection{Variation of the Mean Free Path}
\label{subsect:100_AU}


The second effect of the AE bias is a variation in the mean free
path. Since gravitation is a long-range interaction, there are
three different factors that can  change the CDM particle
velocity. One is the gravitation of the existing halo, i.e., after
the halo has formed. The mean free path of the particles $L$ will
then be about the size of the halo, $R$. Another factor is the
scattering of two CDM particles. Suppose that particle 1 passes by
a stable particle 2 with velocity $v$. If particle 1 gains a
scattering angle larger than $45^{o}$ (in this case $|\delta
v|\sim |v|$, and we call it a 'scattering'), the impact parameter
$b$ should satisfy $b<2Gm/v^2$, where $m$ is the particle mass.
Let $n$ be the particle number density and set the $v^2$ equal to
the velocity dispersion $\sigma_v^2$; then the mean free path of
scattering $L_s$ should satisfy $n (L_s\pi b^2)\geqslant 1$, and
we get
\begin{equation}
\label{effectLs}
L_s=\sigma_v^4/(4\pi G^2 m \rho)
\end{equation}
If we define $T_v \sigma_v = L_s$, we can also expect the two-body
relaxation time $T_v$ to be proportional to $\sigma_v^3/{G^2 m
\rho} \propto 1/m \propto N$ (for a given $\sigma_v$ and density
$\rho$). The third resulting proportionality comes from summing
the velocity deflections from distant particles. Binney \&
Tremaine(1987) estimated the relaxation time that was due to this
cause to be proportional to $N/ln N$. But analytical and
simulation results (Huang et al.1993; Diemand et al.2003) show
that the relaxation time is proportional to $N$, not $N/ln N$, if
the softness factor is taken into account in the simulations. Here
we use eq (\ref{effectLs}) to discuss the effect of the AE bias.

As we can see from the formula above, for certain values of $\rho$
and $\sigma_v$, the AE bias will raise the value of $m$ and make
$L_s$ shorter. For a virialized halo, we can estimate the velocity
dispersion from galaxy dynamics (Binney \&Tremaine,1987);
$\sigma_v^2\approx GM/R$, with $M$ the total mass, and $R$ the
radius of the halo. Substitute $\sigma_v$ into
eq.(\ref{effectLs}), and we have

\begin{equation}
\label{effectL}
L_s=\left(\frac{R}{3}\right)\left(\frac{M}{m}\right)
\left(\frac{\bar{\rho}}{\rho}\right)
\end{equation}
where $\bar{\rho}=M/{{{3}\over{4}} \pi R^3}$. If the AE bias
causes $m\rightarrow m f_{AE}$ and $N\rightarrow N/f_{AE}$, then
we get
\begin{equation}
\label{effectLT}
L_s\rightarrow L_s/f_{AE}\qquad\mbox{ and }\qquad
T_v\rightarrow T_v/f_{AE}
\end{equation}

For example, in the central region of a typical halo in the
simulation, let $R=1Mpc/h$, $M/m=N_{halo}=10^6$, and
$\rho/\bar{\rho}=10^5$, and we then get $L_s=3.3Mpc/h$, close to
the dimension of the halo. If we apply the expression of the
scattering cross section commonly used in the SIDM model,
$\sigma_{SI}=1/(L \rho)$, and if we let the central area density
of the halo be $\rho\approx100Gev/cm^3$, we can estimate
$\sigma_{AE}\approx9\times10^{-26}cm^2/GeV$. In the above, we only
used typical data of the central area of halos. Indeed, the mean
scattering cross section of the whole halo should be smaller.
However, considering that the mass of the DM particle candidates
is on the order of GeV, $f_{AE}\sim 10^{70}$, from
eq.(\ref{effectLT}) we can see that the actual scattering cross
section between two DM particles should be far less than the cross
section in the simulation. In this case, the change of the
particle velocity depends mostly on the deflection by the existing
halo. When discussing the SIDM model, Spergel \&Steinhardt(2000)
and Dave et al.(2001) suggested that if the $\sigma_{SI}$ of the
DM particles is around $10^{-23}$ to $10^{-24} cm^2 GeV^{-1}$, or
their mean free path $L\sim 1kpc - 1Mpc$ is at the solar radius
level, then the halo center cusp problem and the excessive
substructure problem can be solved. Now we can see that the AE
bias has provided an unexpected short mean free path $L_{AE}$ and
a $\sigma_{AE} \lesssim\sigma_{SI}$ smaller than the SIDM model
expectation.

To give a rough estimate here, we set the simulation timescale to
be $t_{all}\sim 100Gyr$, and the particle velocity
$\overline{v}\sim1000km s^{-1}$, and $L_s\sim R\sim 1Mpc$. Then
one particle in simulation will scatter for about
$t_{all}/t_{scatter}=t_{all}/(L_s/\overline{v})\sim 10$ times. Yet
the real CDM particles will $not$ experience such scattering for
they have a far more smaller scattering cross section. This is the
difference caused by the AE bias.

Unfortunately, at present, there is not a quantitative theory on
how the mean free path $L_s$ will affect the history of structure
formation, but the qualitative picture is clear. When $L_s\ll R$,
a short $L_s$ means a large scattering cross section and a high
thermal conductivity, which might alleviate the cusp problem and
make the halos more spherical. But some authors argue that a short
$L_s$  can also drain energy from the halo centers, which could
lead to core collapse (Burkert,2001). When $L_s\gg R$, a large
$L_s$ means a small scattering cross section and a large average
relaxation time $T_v$. Eq.(\ref{effectLT}) shows us that with
improvement in the mass resolution (decrease in $f_{AE}$), $T_v$
will be increased. The simulation by Diemand, et al.(2004) shows
that $T_v\propto N$ and that it can easily rise to more than 1000
Gyr by increasing the total particle number $N$ in the simulation.
In this case, the tidal force is unable to damage small halos in
one Hubble time, as was shown in the simulation by Kazantzidis,et
al.(2003). Here the substructures are mainly what remain of halo
merging. A very small scattering cross section also means that the
effects of the two-body scattering of the CDM particles can be
neglected and that a very poor thermal conductivity is expected.
If $T_v$ is far much longer than one Hubble time, then it is hard
to make the CDM halo virialize with the poor thermal conductivity.
And a new density distribution implies a different particle energy
distribution, so there will not be enough time to form new
substructures on scales less than $R$, such as the $\rho\sim
r^{-2}$ cusp profile in the isothermal model.

The current simulations have brought in an unexpectedly short
$L_s\sim R$, not in both cases above, and it is different from the
real CDM particle systems ($L_s\gg R$). The corresponding
scattering cross section is smaller than what is needed in the
SIDM model, and it can not be neglected. The simulation results
will be sensitive to $L_s$ and to the relaxation effects of
subhalos, such as the number and mass density of subhalos (Ma et
al.2004). One can expect that, with improvements to the numerical
technique, the AE bias will be gradually alleviated, and the
simulation results will gradually move to the $L_s\gg R$ regime.
By increasing $L_s$, the halo center cusp problem can be
alleviated. But the relaxation time will soon rise up to more than
one Hubble time. As a result, halos will tend to be more triaxial,
and more small halos will survive the process of merging. In other
words, the AE bias decreases the number of substructures. The
substructure problem in future simulations can become more
serious. Actually, this is in agreement to a certain extent with
the discussion of gravitational lensing (Mao et al.2004) in which
more substructures are expected in real DM halos than in the
current simulations. But if $L_s$ increases more, the thermal
conductivity in the halos will become poor, which means it will be
harder to form enough virialized halos in one Hubble time.

By the way, eq.(\ref{effectL}) can help us understand the
concentration properties of DM halos in simulations. In a given
simulation in which the particles have a fixed mass $m$, bigger
halos will have larger total masses $M$. Then, based on
eq.(\ref{effectL}), one can predict that the big halos will have
larger $L_s/R$ than small halos. Large $L_s/R$ mean that the
particles can move more freely in the halo, which will become more
dispersed, so equation(3) can give a rational explanation of why
the more massive halos in the simulations have a smaller
concentration coefficient $C$. Eq.(\ref{effectL}) is available for
all virialized particle systems with a pure Newton gravitation
interaction. If we suppose the particles to be the stars of
elliptical galaxies, then the discussion above will predict that
the more massive elliptical galaxies will be more dispersed. In
other words, they will have a fainter mean surface brightness, and
this is consistent with the observations (Binney \&Merrifield,
1998,p.205).



\section{Conclusion and Discussion}
\label{sect:disc}

In summary, we have pointed out that a physical bias exists in
current cosmological simulations. We find that the AE bias can
change the small-scale properties of the CDM particle system in
two respects. (1) An upper limit is imposed on the spatial
resolution of the simulation results. (2) An unexpected short mean
free path of the system is generated. The corresponding two-body
scattering cross section is near the value expected in the SIDM
model.

What can we learn from the discussion above? The upper limit
$L_{min2}$ of the first effect can help us estimate the reliable
range of the simulation results. The second effect can alter the
simulation results on small scales and cause the halo center to be
more cuspy and more spherical. These results can help us
understand the "cusp problem" and the "substructure problem" in
the current simulations, and they can help us reconstruct a
correct scenario for the clustering and evolution of the real dark
matter in the universe. Because the AE bias factor $f_{AE}$ is
currently about $10^{70}$, the mean free path $L_s$ and the
relaxation time $T_v$ of real CDM particles systems should be much
larger than we see in the simulations. When $L_s$ is far more
larger than the halo diameter, the halo will be dispersed and will
be hard put to form a cuspy density profile. When $T_v$ is much
longer than one Hubble time, CDM halos will not have enough time
to be virialized of the poor thermal conductivity (this is
contrary to the prevailing view of the popular halo model), and
nearly all the centers of the existing halos can survive the
process of merging until the present time. Our results imply (1)
that real CDM halos may not have the cusp problem(the cusps of the
simulation halos are mainly the effect of the AE bias)and (2) that
real CDM halos may not have enough time to be virialized. This
result can help us understand the substructure problem. The
discussion above also tells us that the theoretical analysis of
the halo property, which is based on plenty of thermal
conductivity, is available for simulation halos but not for the
real CDM halos.

On the other hand, since it is hard to form new virialized halos
in one Hubble time and since most existing small halos (coming
from primary disturbances) can survive the process of merging, the
number of these small halos may retain some information about the
primordial disturbance. If we believe the number of dwarf galaxies
is proportional to the number of small halos, then the percentage
of dwarf galaxies in galaxy clusters can tell us something about
the small-scale properties of the primordial disturbance.

{\bf Acknowledgements:} We thank Professor T. Kiang, Alessandro
Romeo and Dr. Steen Hansen for checking the manuscript and for
helpful suggestions.


\clearpage

\end{document}